\documentclass[sigconf]{acmart}
\renewcommand\footnotetextcopyrightpermission[1]{} 
\setcopyright{none}
\usepackage{amsmath,amssymb,amsfonts}
\usepackage{algorithmic}
\usepackage{graphicx}
\usepackage{textcomp}
\usepackage{xcolor}
\usepackage{subfigure}

\def\BibTeX{{\rm B\kern-.05em{\sc i\kern-.025em b}\kern-.08em
    T\kern-.1667em\lower.7ex\hbox{E}\kern-.125emX}}

\acmConference[]{}{}{}

\begin{document}

\title{Android Malware Detection based on Factorization Machine}

\author{Chenglin Li}
\affiliation{%
	\department{Dept. Electrical and Computer Enginerring}
	\institution{University of Alberta}
	\city{Edmonton}
	\state{Alberta}
	\country{Canada}
}
\email{ch11@ualberta.ca}

\author{Keith Mills}
\affiliation{%
	\department{Dept. Electrical and Computer Enginerring}
	\institution{University of Alberta}
	\state{Alberta}
	\country{Canada}
}
\email{kgmills@ualberta.ca}

\author{Di Niu}
\affiliation{%
	\department{Dept. Electrical and Computer Enginerring}
	\institution{University of Alberta}
	\city{Edmonton}
	\state{Alberta}
	\country{Canada}
}
\email{dniu@ualberta.ca}

\author{Rui Zhu}
\affiliation{%
	\department{Dept. Electrical and Computer Enginerring}
	\institution{University of Alberta}
	\city{Edmonton}
	\state{Alberta}
	\country{Canada}
}
\email{rzhu3@ualberta.ca}

\author{Hongwen Zhang}
\affiliation{%
	\department{CEO \& CTO}
	\institution{Wedge Networks}
	\city{Calgary}
	\state{Alberta}
	\country{Canada}
}
\email{hongwen.zhang@wedgenetworks.com}

\author{Husam Kinawi}
\affiliation{%
	\department{President}
	\institution{Wedge Networks}
	\city{Calgary}
	\state{Alberta}
	\country{Canada}
}
\email{husam.kinawi@wedgenetworks.com}

\begin{abstract}
As the popularity of Android smart phones has increased in recent years, so too has the number of malicious applications. Due to the potential for data theft mobile phone users face, the detection of malware on Android devices has become an increasingly important issue in cyber security. Traditional methods like signature-based routines are unable to protect users from the ever-increasing sophistication and rapid behavior changes in new types of Android malware. Therefore, a great deal of effort has been made recently to use machine learning models and methods to characterize and generalize the malicious behavior patterns of mobile apps for malware detection. 

In this paper, we propose a novel and highly reliable classifier for Android Malware detection based on a Factorization Machine architecture and the extraction of Android app features from manifest files and source code. Our results indicate that the numerical feature representation of an app typically results in a long and highly sparse vector and that the interactions among different features are critical to revealing malicious behavior patterns. After performing an extensive performance evaluation, our proposed method achieved a test result of 100.00\% precision score on the DREBIN dataset and 99.22\% precision score with only 1.10\% false positive rate on the AMD dataset. These metrics match the performance of state-of-the-art machine-learning-based Android malware detection methods and several commercial antivirus engines with the benefit of training up to 50 times faster.
\end{abstract}


\maketitle


\section{Introduction}
\label{sec:intro}


Smartphone usage is prevalent in our daily lives. According to surveys on global OS market shares \cite{androidDevice18, androidmarket18}, Android is the visibly dominant mobile OS with a solid hold of around $75\%$ market share across \emph{all} mobile devices and $85.1\%$ dominance for smartphones specifically in 2018.
The rapid growth of mobile device usage, coupled with the majority market share that the Android OS enjoys have not only brought about opportunities for Android app development, but also serve to emphasize the challenge involved in defending devices from malware. According to Kaspersky's Mobile Malware Evolution Reports for 2016 through 2018 \cite{RomanUnuchek2016, RomanUnuchek2017,VictorChebyshev2018}, the number of malicious installation packages amounted to $8,526,221$, $5,730,916$ and $5,321,142$, respectively. While these numbers indicate a downward trend, one should not be fooled as the number of new Trojans targeting financial information was $128,886$, $94,368$, and $151,359$ each year, respectively, indicating that these classes of malicious, theft-enabling software constitute a larger proportion of Android malware each year - from $1.51\% $ in 2016 to $1.65\% $ in 2017 to an alarming $2.84\% $ in 2018.

To win the battle and protect mobile phone users, a number of anti-virus companies, like McAfee and Symantec, provide software products as a major defense against these kinds of threats. These products typically use a signature-based method \cite{venugopal2008efficient} to recognize threats. 
Signature-based methods involve the generation of a unique signature for each previously known malware, while detection involves scanning an app to match existing signatures in a malware database. 
On the other hand, the heuristic-based method, introduced in the late 1990s, relies upon explicit expert rules to distinguish malware, giving rise to errors induced by human bias. In fact, both methods will be less effective if the development of the malware database or expert rules cannot keep pace with the speed at which new malware emerges and evolves.


To overcome the aforementioned problems, an alternative emerging approach is to develop intelligent malware detection techniques based on Machine Learning (ML), whose generalization capabilities are capable of discovering unintuitive patterns in previously undetected malware samples. One major type of machine learning-based malware detection method is called static analysis \cite{wu2012droidmat, arp2014drebin}, which can make decisions about an app without executing it in a sandbox, thus incurring a low overhead for execution. 
Static analysis has two phases: feature extraction and classification. In the first phase, various features such as API calls and binary strings are extracted from an original file. In the second phase, an ML model learns to automatically categorize the file sample into malware or benign-ware based on a vectorized representation of the file. 
For example, DroidMat \cite{wu2012droidmat} performs static analysis on the manifest file and the source code of an Android app to extract multiple features, including permissions, hardware resources, and API calls. It then uses $k$-means clustering and $k$ nearest neighbor ($k$-NN) classification to detect malware. 
DREBIN \cite{arp2014drebin} extracts similar features from the manifest file and source code of an app and uses a Support Vector Machine (SVM) for malware classification based on one-hot encoded feature vectors.

However, existing machine learning techniques for malware detection have yielded limited accuracy with high false positive rates, mainly due to the use of first-order models or linear classifiers, such as SVM \cite{arp2014drebin}. These are insufficient to discover all malicious patterns. A natural idea to introduce non-linearity into malware detection is to consider the interaction between features, or in other words, \emph{feature crossing} or \emph{basis expansion}. For example, an app concurrently requesting both GPS and \texttt{SEND\_SMS} permissions may be attempting to execute a location leakage, while the presence of either one of such requests alone does not point to any malicious behavior. However, ML models involving feature crossing are not scalable to long feature vectors.


For example, a total number of $545,000$ features are used by DREBIN \cite{arp2014drebin}, the SVM-based detector, which means that more than $297$ billion interactions need to be considered if feature crossing was used. One could expect this number to be even larger in a more recent dataset; the Android Malware Dataset (AMD) \cite{wei2017deep}, which contains more file samples thus exposing more features. Moreover, although the total number of features is large, the number of features activated by each file sample is usually much smaller, leading to a sparse vectorized representation for each individual app. This will further lead to even sparser interaction terms (the crossed terms), posing significant challenges to model training---there are not enough non-zero entries in the dataset to train the coefficient of each crossed term.

The goal of this paper is to accurately model feature interactions and efficiently handle long and sparse features. To this end we propose a novel \emph{Factorization Machine} (FM) model for Android malware detection. In contrast to feature crossing or basis expansion, which suffers from the model size issue and the sparsity issue mentioned above, Factorization Machines \cite{rendle2010factorization} aim to learn the coefficient of each interaction as the inner product of two latent vectors, thus effectively reducing the number of parameters to linear to $n$, where $n$ is the length of the feature vector.

We evaluated our model on two Android malware datasets: DREBIN \cite{arp2014drebin} and AMD\cite{wei2017deep}, which contain 5560 and 24553 samples, respectively. In order go gauge performance, the metrics we utilized consisted of accuracy, false positive rate (FPR), precision, recall and F1 
In addition, we also evaluated the performance when identifying specific families of malware, which is especially important given the growing share of banking Trojans. With respect to this task, we focused primarily on accuracy and false positive metrics. 



The remainder of this paper is organized as follows: Section~\ref{sec:prelim} reviews the background of the Android system and our feature extraction technique, while Section~\ref{sec:fm} describes the mathematics behind a Factorization Machine. In Section~\ref{sec:simu} we elaborate on the two datasets used in this experiment, formally describe the metrics we are using before stating our test results in Section~\ref{ssec:detect} and interpret them in Section~\ref{ssec:interpretFM_MLP}. Next, we compare our test results to several several popular antivirus engines and gauge our model's detection rating with respect to specific malware families in Sections~\ref{ssec:commerical} and \ref{ssec:family}. Finally, we discuss future work and list a few related research projects in Sections~\ref{sec:future} and \ref{sec:related}, respectively, before concluding in Section~\ref{sec:conclude}. 

\section{Android Feature Extraction}
\label{sec:prelim}

Android applications are written in Java and executed within a custom Java Virtual Machine (JVM). Each application package is contained in a jar file with the extension of \texttt{apk}. Android applications consist of many components of various types, which are the essential building blocks of the application. Each component has an entry point through which the system or a user can enter the application. In addition, there are four fundamental building blocks of an Android app: \emph{Activities}, \emph{Services}, \emph{Broadcast Receivers} and \emph{Content Providers}. All components must be declared in the application manifest file in order to be used. Communication between these components is achieved by using intents and intent filters. Intents are messaging objects that can be used to request actions from other application components while intent filters are expressions declared in the application manifest file that specify the intent type that a component will receive. Since application components interact via the intent method, it is critical to analyze both the components themselves, as well as their communication intents, for security concerns.

Before classification on any model can be done, raw data must be processed. The feature engineering section of our malware detection system consists of three parts: Unpacking and Decompiling, Feature Extraction, and Encoding -- all shown below in Fig ~\ref{fig:sys}. 
\begin{figure*}
	\centering
	\includegraphics[width=6.8in]{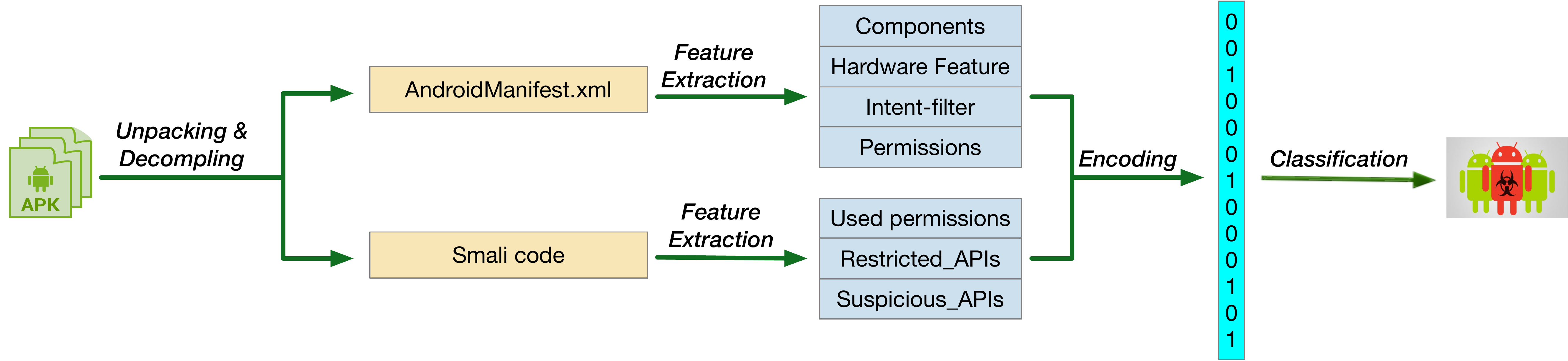}
	\caption{System architecture of our Malware detection model.}
	\label{fig:sys}
\end{figure*}

\textbf{Unpacking and Decompiling}:
Each \texttt{apk} file is actually a specialized zipped file that consists of the application source code, resources, assets, and manifest file. The source code is encoded as \texttt{dex} files (i.e., Dalvik Executable Files) that can be interpreted by the Dalvik VM. The manifest file consists of a number of declarations and specifications. Finally, other resources may contain images, HTML files, etc.. Since the \texttt{dex} files are compiled, binary executable code, and therefore not meant to be read or interpreted, features cannot be readily extracted from them directly. Therefore, they must be decompiled into other formats that can be read and interpeted, such as Smali code or even Java code. Smali code is an intermediate form that is decompiled from the dex files; it is essentially the assembly code format of an application. Only after then \texttt{apk} files have been decompiled can we continue onto our next step.


\textbf{Feature extraction}: Feature extraction is one of the the most important aspects involved in the training of a machine learning model. The upper bound of a given model's performance directly depends on the nature of the features used. After performing a study of the Android system and comparing it to previous work experience in its field, we chose to extract $7$ kind of features from both the source code and manifest file. The following four types of features are extracted from a given app's manifest file:
\begin{enumerate}
\item \textbf{App components}: The components declared in the manifest file define the different interfaces that exist between the app and the end-user and the app and the larger Android OS as a whole. The names of these components are collected to help identify variants of well-known malware, for example the DroidKungFu family share the name of several particular services \cite{arp2014drebin}. 
\item \textbf{Hardware features}: If an application wants to request access to the hardware components of the device, such as its camera, GPS or sensors, then those features must be declared in the manifest file. Requesting certain hardware components may have security implications, For example, requesting usage of the GPS and network modules may be a sign of location leakage. 
\item \textbf{Permissions}: Android uses a permission mechanism to protect the privacy of users. An app must request permission to access sensitive data (e.g. SMS), system features (e.g. camera) and restricted APIs. Malware usually tends to request a specific set of permissions. In this respect, this is similar to how we handle hardware features. 
\item \textbf{Intent filter}: Intent filters declared within the declaration of components in the manifest file are important tools for inter-component and inter-application communication. Intent filters define a special entry point for a component as well as the application. Intent filters can be used for eavesdropping specific intents. Malware is sensitive to a special set of system events. Thus, intent filters can serve as vital features.
\end{enumerate}

Furthermore, we also extract another three types of features from the decompiled application source code (e.g., Smali code): 
\begin{enumerate}
\item \textbf{Restricted APIs}: In the Android system, some special APIs related to sensitive data access are protected by permissions. If an app calls these APIs without requesting corresponding permissions, it may be a sign of root exploits.
\item \textbf{Suspicious APIs}: We should be aware of a special set of APIs that can lead to malicious behavior without requesting permissions. For example, cryptography functions in the Java library and some math functions need no permission to be used. However, these functions can be used by malware for code obfuscation. Thus, attention should be paid to the unusual usage of these functions. We mark these types of functions as \emph{suspicious APIs}. 
\item \textbf{Used permissions}: We first extract all API calls from the app source code, and use this to build a set of permissions that are actually used in the app by looking up a predefined dictionary that links an API to its required permission(s). 
\end{enumerate}

\textbf{Encoding}:
Next, we encode our extracted features into a common format that can be fed into any generic classifier. Our method uses an $N$-dimensional indicator to encode each application into a feature representation, where $N$ is the feature dimension. To be specific, suppose all the extracted features form a feature set $S$ with size $|S|$, 
then we represent each \texttt{apk} file as a binary vector of length $|S|$, whose entries are $1$ only if a given feature is used by the app.


\begin{figure}
	\centering
	\includegraphics[width=3.3in]{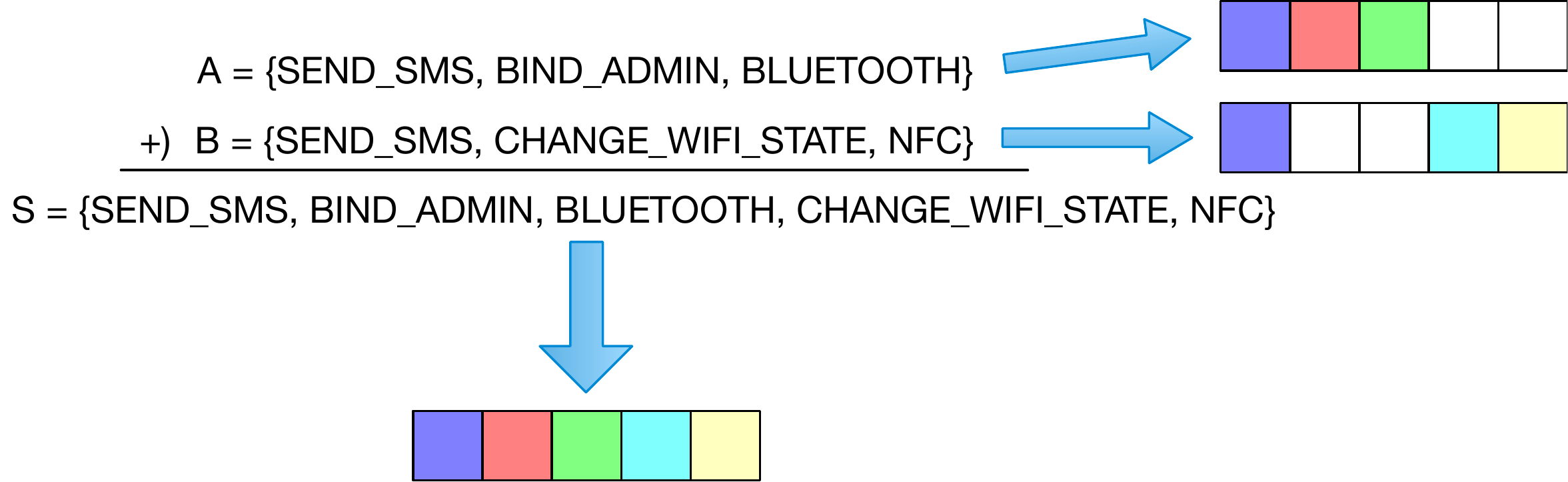}
	\caption{One-hot encoding for string features.}
	\label{fig:one_hot}
\end{figure}

For example, suppose we have two Android applications, A and B, which each request three permissions as illustrated in Fig.~\ref{fig:one_hot}\footnote{Here we use different color blocks to represent different feature values in permission set, although in practice active features are represented by a '1'. White blocks mean the feature is not used (set to '0').}. As there are five unique permissions requested by A and B, we can then create a vector $\mathbf{x}_A, \mathbf{x}_B \in \{0,1\}^5$ such that each entry represents exactly one permission, e.g., the first entry as a blue block represents the permission \texttt{SEND\_MSG} and the second entry represents the permission \texttt{BIND\_ADMIN}. As a result, we can write $\mathbf{x}_A = (1,1,1,0,0)$ and $\mathbf{x}_B=(1,0,0,1,1)$. It is straightforward to extend this idea to all kinds of extracted features as discussed in Sec. 2. The formal name for this scheme in literature is \emph{one-hot} encoding.

\begin{figure}
	\centering
	\includegraphics[width=2.7in]{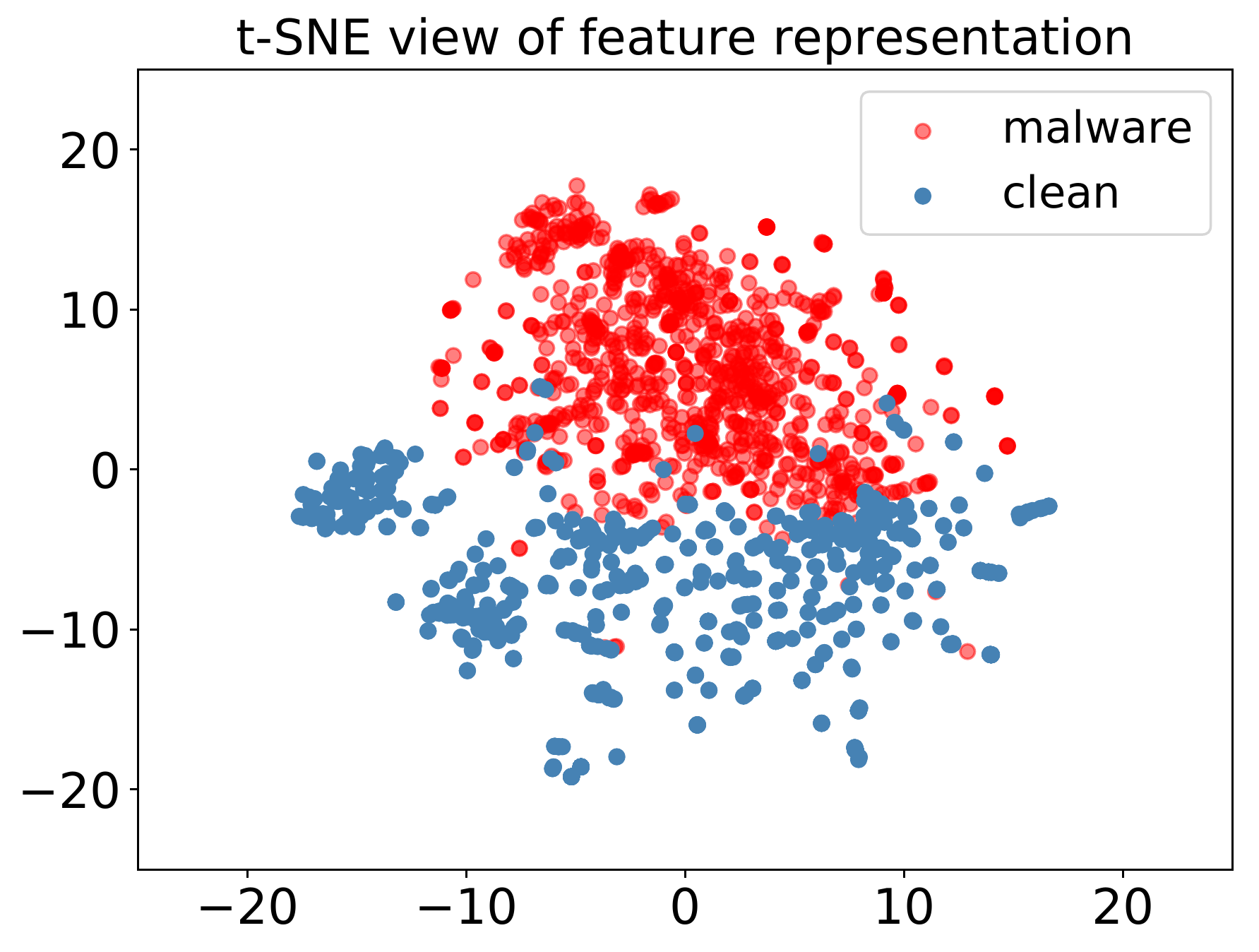}
	\caption{t-SNE view of $2000$ samples.}
	\label{fig:tsnedrebin}
\end{figure}

To visualize the effectiveness of our feature extraction and representation technique in distinguishing between malware and clean files, we applied a t-SNE \cite{maaten2008visualizing} algorithm on $2000$ already encoded samples, $1,000$ of which came from the DREBIN \cite{arp2014drebin} dataset and while $1,000$ were clean. The result is shown in Fig~\ref{fig:tsnedrebin}. Through this representation we can clearly see that the malware and clean files have formed several visibly identifiable, yet overlapping clusters, which implies the need  for a non-linear classifier in order to accurately discriminate each class.


\section{Factorization Machine for Malware Detection}
\label{sec:fm}

Generally, a classification problem in machine learning is to infer a function $h: \mathbb{R}^{n} \to \mathbb{R}$ for all possible $\mathbf{x} \in \mathbb{R}^n$ to predict how much it belongs to a class. To find such a function, we are given a set of samples, each of which has been marked as a ``malicious'' or ``benign''.

In an abstract sense, the goal at the core of this paper is one that involves achieving a high accuracy on a binary classification problem using a classifier that is not very well-known when compared to the peers its performance is contrasted against. Specifically, we have chosen to use a Factorization Machine~\cite{rendle2010factorization} to meet this objective, for reasons to be explained below.


\subsection{Limitations of First-Order Classifiers}
\label{ssec:FirstOrder}

Given a data sample $ \vec{x} $, a typical machine learning algorithm will attempt to determine its class, $ \hat{y}(\vec{x}) $\footnote{This is usually a probability (binary) or vector of class probabilities (multiclass) that are processed later}, by learning a set of weights, $ \vec{w} $, such that,

\begin{equation}
\centering
\label{eq:ML}
\hat{y}(\vec{x}) = h(\vec{x}; \vec{w})
\end{equation}

Where \emph{h} is sometimes known as the \emph{transfer function} of the algorithm. For example, in the case of Support Vector Machines, used by DREBIN~\cite{arp2014drebin}, \emph{h} can be written as,

\begin{equation}
\centering
\label{eq:SVM}
	h(\vec{x}; \vec{w}) = \vec{x}^T\vec{w} + w\textsubscript{0}
\end{equation}

Where $ \vec{w}\textsubscript{0} $ is the intercept coefficient, technically a part of $ \vec{w} $ but always multiplied by 1.
Most of the basic, well-known classifiers, including Support Vector Machines,  Naive Bayes (NB), and Logistic Regression, operate in a similar manner, where the input sample is compared to the learned weights allowing a class decision to then be made.

These models are not suitable for Android malware detection for two reasons: \emph{First}, as Figure~\ref{fig:one_hot} implies, the feature vectors from one-hot encoding consist of ones and zeroes and are likely to be highly sparse. For example, samples in the benchmark dataset DREBIN~\cite{arp2014drebin} will be encoded into vectors with $93,324$ entries, and on average only $73$ features are non-zero, which makes weight training difficult, because a weight value of $55$ is as accurately as a weight value of $-249$ if that weight is always multiplied by a $0$. 

\emph{Secondly}, these models only exploit the First-Order information found within the features -- interactions between features and weights. They do not take interactions \emph{among} the features themselves into account. For example, going back to Figure~\ref{fig:one_hot} again, if a specific class of Malware can be reliably detected by checking if it requests a certain set of features (e.g. \texttt{BLUETOOTH} and \texttt{CHANGE\_WIFI\_STATE}) together - meaning it requests all of them not just a few - then a good starting point for a reliable classifier is one that can detect if that specific set of features are active. This is a second-order interaction - first-order classifiers such as SVMs and Naive Bayes cannot handle these automatically unless an interaction term between these features was added previously in the feature engineering stage. The inclusion of these interaction terms requires \emph{a priori} knowledge regarding the malware and also serve to expand the number of features. 


\subsection{Second-Order Feature Crossing and Factorization Machines}
\label{ssec:SecondOrder}

Typically Multi-Layer Perceptrons (MLP) and Deep Neural Networks (DNN) are the go-to solution for solving hard classification problems due to the properties they possess as universal function approximators~\cite{ruck1990multilayer, SCARSELLI199815}. However, the number of parameters involved in training these models for a given task involve the tuning of very complex model with a large number of synaptic weights, even before feature interactions are introduced. With that in mind, consider a natural method for learning interactions of different features is through basis expansion or feature-crossing:
\begin{equation}
h(\mathbf{x}) = w_0 + \sum_{i=1}^{n} w_{i} x_i + \sum_{i=1}^{n}\sum_{j=i+1}^{n}W_{ij} x_i x_j.
\label{eq:reg}
\end{equation}

By assigning a weight $W_{i,j}$ for each pair of $x_i$ and $x_j$, we have the easiest way to capture pairwise interactions. However, it is not efficient here due to the large number of parameters: this model has $n(n-1)/2$ free parameters. As stated in Section~\ref{ssec:FirstOrder}, the input vector for the DREBIN~\cite{arp2014drebin} dataset has a length of $93,324$ but the number of nonzero entries is about $73$ on average. In this case, full feature crossing like $W$ would necessitate roughly four billion weights! This brings heavy burdens on the training process since the model becomes so complicated it requires large computational resources and is very time-consuming. This problem is further compounded by the sparse nature of the data in this problem, which makes the task of applying meaningful updates to a large number of weights very difficult. Hence, MLPs are not an optimal solution to this problem.

\begin{figure}
	\centering
	\includegraphics[width=2.0in]{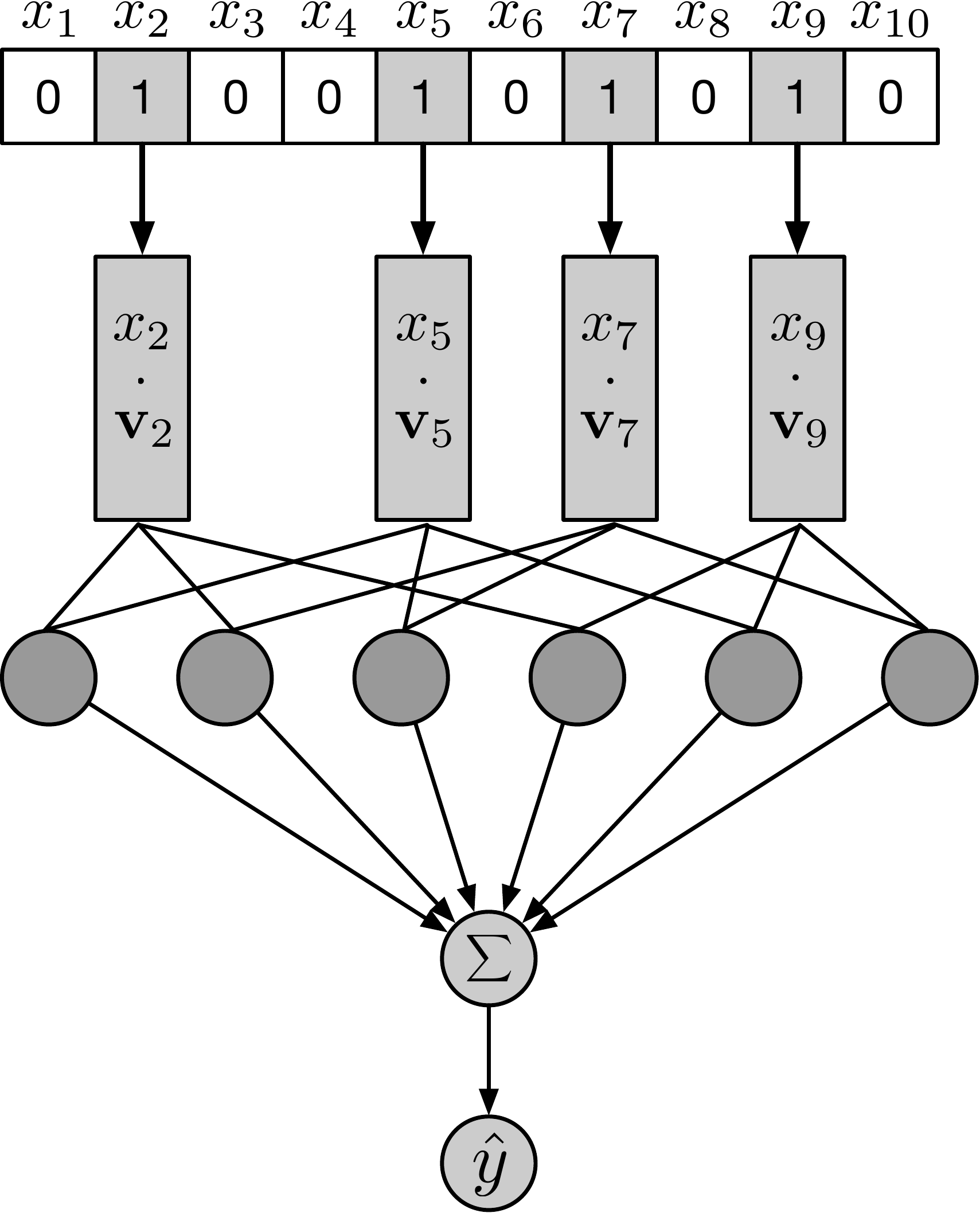}
	\caption{The architecture of Factorization Machine model.}
	\label{fig:fm}
\end{figure}

Popular techniques to overcome the issues mentioned above and in Section~\ref{ssec:FirstOrder} are low-rank or dimension reduction methods. In particular, we have chosen to use a classifier that implements feature interactions in the \emph{learning stage} - the Factorization Machine~\cite{rendle2010factorization} (FM), described by Figure~\ref{fig:fm}\footnote{The dark gray node stands for the inner product operator.}. More specifically, FM assumes that $W$ is with the largest rank of $k$ and therefore, we can decompose $W = V V^{\sf T}$. 

If we denote $\mathbf{v}_i$ as the $i$-th row of $V$, FM will train a hidden vector $\mathbf{v}_i$ for each $x_i$ and models the pairwise interaction weight $w_{ij}$ as the inner product of the corresponding hidden vectors of entries $x_i$ and $x_j$:
\begin{equation}\label{eq:FM}
h(\vec{x}) = w_0 + \sum_{i=1}^{n} w_i x_i + \sum_{i=1}^{n}\sum_{j=i+1}^{n}\langle \mathbf{v}_i,\mathbf{v}_j \rangle x_ix_j,
\end{equation}
where $\langle \cdotp,\cdotp \rangle$ denotes the dot product of two vectors of length $k$:
\begin{equation}\label{eq:parameters}
\langle \mathbf{v}_i,\mathbf{v}_j \rangle := \sum_{f=1}^{k}v_{i,f} v_{j,f},
\end{equation}

The parameters that are learned during the training stage -- $ w\textsubscript{0}, w\textsubscript{i}, v\textsubscript{i} $ and $ v\textsubscript{j} $ -- can be updated using generic stochastic methods such as gradient descent~\cite{rendle2010factorization, yaochenAndDiFM} with augmentations such as Adam~\cite{adamOptimizer}. 
In practice, the hyperparameter $k$ is much smaller than the feature dimension $n$ ($k \ll n$). Thus, the number of parameters to be estimated reduces from $O(n^2)$ to $O(nk)$.

We can further improve the performance of FM by using more sophisticated feature engineering schemes for cross terms. For example, by using ``partial FM'', which only involves interactions between selected features, e.g., between \texttt{Used permissions} and \texttt{Permissions}, thus ignoring crossed terms that are not relevant to malicious behavior discovery.

\section{Experiments}
\label{sec:simu}

In this section, we evaluate the performance of our Factorization Machine-based Android malware detection system. We apply our system to malware detection tasks and malware family identification tasks, based on two public benchmark datasets: DREBIN \cite{arp2014drebin} and AMD \cite{wei2017deep}. We also check our FM model against popular antivirus engines and state the logistics of our decompilation and feature extraction procedure.


\begin{table}[]
\centering
\caption{Performance metrics of Android malware detection.}
\label{table:metrics}
\begin{tabular}{ll}
\hline
Metrics   & Description                               \\ \hline
$TP$        & \# of malicious apps correctly detected \\
$TN$        & \# of benign apps correctly classified  \\
$FP$        & \# of false prediction as malicious     \\
$FN$        & \# of false prediction as clean         \\
$Accuracy$  & $ (TP + TN) / (TP + TN + FP + FN) $	  \\
$Precision$ & $TP/(TP+FP)$                            \\
$Recall$    & $TP/(TP+FN)$                            \\
$F1$        & $2*Precision*Recall/(Precision+Recall)$ \\
$FPR$ 		& $FP/(FP + TN) $ \\
\hline
\end{tabular}
\end{table}

\subsection{Android \texttt{APK} Data}
\label{ssec:setup}
To perform this experiment, we used two public benchmark datasets: DREBIN~\cite{arp2014drebin}, which has been mentioned previously, and AMD~\cite{wei2017deep}:
\label{exp:datasets}

\begin{itemize}
	\item \textbf{DREBIN}: Contains $5560$ malware files collected from August 2010 to October 2012. All malware samples are labeled as 1 of 179 malware families. This is one of the most popular benchmark datasets for Android malware detection.
	\item \textbf{AMD}: Known as the Android Malware Dataset, it contains $24553$ samples that are categorized in 135 varieties among 71 malware families. This dataset consists of samples collected from 2010 to 2016. This is one of the largest, public datasets. AMD provides more recent evolution trends for Android malware when compared to DREBIN.
\end{itemize}

Further details regarding these two datasets are shown in Table~\ref{table:experimentsdata}. When doing experiments on the AMD dataset, we evaluated all $16753$ clean files. When evaluating on the DREBIN dataset, we randomly sampled $5600$ clean files to match the number of malware samples in this dataset. To simplify our terminologies, the DREBIN dataset (or the AMD dataset) consists of both clean samples and malware samples in the subsequent to this section. 
Also from Table~\ref{table:experimentsdata} we see the overall feature set size grows from $93324$ to $294,019$ as the dataset size grows from $11160$ to $41306$. 

We also collected a number of real-world Android applications from the internet. Resources of these files include \texttt{apkpure} \cite{Apkpure} with 5400 samples, 700 samples from \texttt{360.com} and 13K commercial applications from the HKUST Wake Lock Misuse Detection Project \cite{liu2016understanding}. In total, we have $19100$ real-world applications. 
Then, we uploaded all these files to VirusTotal, a public anti-virus service with 78 popular engines, and inspected scanning reports for each file as a check to ensure that they were truly clean files. Each engine in VirusTotal would show one of three detection results: \texttt{True} for ``malicious'', \texttt{False} for ``clean'', and \texttt{NK} for ``not known'', respectively. If an application had more than one \texttt{True} result, we labeled it as \texttt{malware}; otherwise, we considered it as \texttt{clean}. Thus, only $16753$ out of $19$K collected samples are labeled as \texttt{clean}, and we will only use these samples in further experiments.

In our system, we used APKtool\cite{apkTool} to decompile the source code into Smali code and extract information from the \texttt{AndroidManifest.xml} file. We found this procedure to be quite time-consuming. However, for different applications this would often take a fixed processing time due to the fixed feature space size. Therefore, we focused on evaluating the processing time for unpacking, decompiling and feature extraction, then give out an average processing time for all applications on the encoding and prediction phase.

\subsection{Classifier Evaluation}
\label{ssec:Eval}

\begin{table}[]
	\centering
	\caption{Datasets for detection performance evaluation.}
	\label{table:experimentsdata}
	\begin{tabular}{lllll}
		\hline
		Dataset   & \# Malware  & \# Clean files   & Total 		& Feature size	\\ \hline
		DREBIN    & 5,560        &     5,600       &  11,160	&	93,324		\\
		AMD       & 24,553 		 &	  16,753	   &  41,306	&	294,019		\\
		\hline
	\end{tabular}
\end{table}

We first evaluated our proposed FM-based method and compared it with other existing baseline algorithms, including SVM\footnote{With a Logistic Kernel}, which is used in DREBIN \cite{arp2014drebin} to achieve a detection rate of $93.9\%$, classical machine learning algorithms such as Naive Bayes using Gaussian, Bernoulli and Multinomial Kernels \cite{wu2012droidmat} and shallow, one-hidden layer neural networks\cite{ruck1990multilayer}.\footnote{Throughout this paper, we use some abbreviations to denote these baseline algorithms for figures and tables. In particular, the name of Algorithm ``NB-Bernoulli'' (NB-B) refers to the Naive Bayes classifier using Bernoulli kernel, and the same for ``NB-Multinomial'' (NB-M) and ``NB-Gaussian'' (NB-G).}  In addition, we also sent all samples, including malware samples, to the VirusTotal service and compared it with commercial anti-virus engines.


	

The dataset was randomly split into training ($80\%$) and testing ($20\%$) sets for both experiments in accordance with the \emph{pareto principle}. All models were trained using stratified 5-fold cross validation for hyper parameter tuning and then tested for performance evaluation. The hyperparameters turned for baseline algorithms and our proposed method were trained and tested in the same manner. Finally, we recorded the time it took to re-train the best model chosen by cross-validation over the entire training set. With exception to our FM, all models were training using Sci-kit Learn's~\cite{scikit-learn} API while we used Polylearn~\cite{polylearn} to train our FM.

Moreover, the metrics we utilized for performance evaluation are listed in Table~\ref{table:metrics}. Specifically, we focused on  \emph{precision}, \emph{recall}, \emph{F1} and \emph{False Positive Rate} (FPR). \footnote{Note that in the literature, recall and false positive rate corresponds to malware detection rate and false alarm rate for the detection system.} The Factorization Machine, Multi-Layer Perceptron and Naive Bayes models all produce \emph{probabilities} that a given sample is malware. If this probability was greater than a certain threshold, $0.5$ in this experiment, it was classified as malware for the purposes of cross-validation and out-of-sample test results.

\subsubsection{Hyperparameter Specifications}
\label{ssec:detect}

\begin{table}
	\centering
	\caption{DREBIN Test Results; Threshold = $0.5$} 
	\label{table:drebin}
	\begin{tabular}{c|cccccc}
		\hline
		Algorithm      & SVM   & NB-G   & NB-B           & NB-M     & MLP            & FM     \\ \hline
		Accuracy (\%)  & 95.65 & 97.72  & 94.71          & 94.71    & \textbf{99.73} & 99.46           \\
		Precision (\%) & 96.34 & 96.25  & 90.25          & 90.86    & 99.91          & \textbf{100.00}     \\
		Recall (\%)    & 94.87 & 99.28  & \textbf{99.82} & 99.37    & 99.55          & 98.92         \\
		F1 (\%)        & 95.60 & 97.74  & 94.95          & 94.93    & \textbf{99.73} & 99.46        \\
		FPR (\%)       & 3.57  & 3.84   & 10.35          & 9.90     & 0.09           & \textbf{0.00}        \\
		\hline
	\end{tabular}
\end{table}

\begin{table}
	\centering
	\caption{AMD Test Results; Threshold = $0.5$}
	\label{table:amd}
	\begin{tabular}{c|cccccc}
		\hline
		Algorithm      	& SVM   & NB-G 	& NB-B   	    & NB-M 		& MLP   & FM             \\ \hline
		Accuracy (\%)   & 92.69 & 93.53 & 90.42         & 86.24     & \textbf{99.05} & \textbf{99.05}            \\
		Precision (\%)  & 93.87 & 90.39	& 86.35    	    & 81.35		&\textbf{99.24}& 99.22 	               \\
		Recall (\%)     & 93.65 & 99.56 & \textbf{99.60}& 99.32 	& 99.14 & 99.16 			   \\
		F1 (\%)         & 93.76 & 94.75 & 92.51 	    & 89.44 	& \textbf{99.19} & \textbf{99.19} 	       \\
		FPR (\%)        & 8.69  & 15.04 & 22.36		    & 32.36 	& \textbf{1.07}  & 1.10  	     \\
		\hline
	\end{tabular}
\end{table}

The best hyperparameters for each algorithm we used -- except for Naive Bayes which had no hyperparameters for cross validation to tune -- are as follows: SVM with a Logistic Kernel preferred a penalty of $1$ and kernel coefficient, $ \gamma $ of 5e\textsuperscript{-5} when the malware and benign classes were weighted equally. The best MLP performance was found by updating the weights in accordance with the Adam~\cite{adamOptimizer} learning rule, the ReLU~\cite{ReLUHinton} activation function, with hidden layers that consisted of 150 and 200 neurons for the DREBIN and AMD sets, respectively with a batch size of 200. For our FM models we found a value of \emph{k} = 10 to be the most optimal. Finally, both the MLP and FM models were trained for 200 epochs.


\begin{table}
	\centering
	\caption{Out-of-Sample Training Times; H:MM:SS Format}
	\label{table:trainTime}
	\begin{tabular}{c|cccccc}
		\hline
		Algorithm & SVM      & NB-G 	& NB-B  & NB-M	& MLP     & FM    	      \\ \hline
		DREBIN    & 0:01:40  & 0:00:13  & 12ms  & 9ms   & 0:09:01 & 0:00:35          \\
		AMD       & 0:21:41  & 0:02:38  & 54ms  & 34ms	& 2:13:42 & 0:02:40          \\
		\hline
	\end{tabular}
\end{table}

\begin{figure}
	\centering
	\includegraphics[width=3.5in]{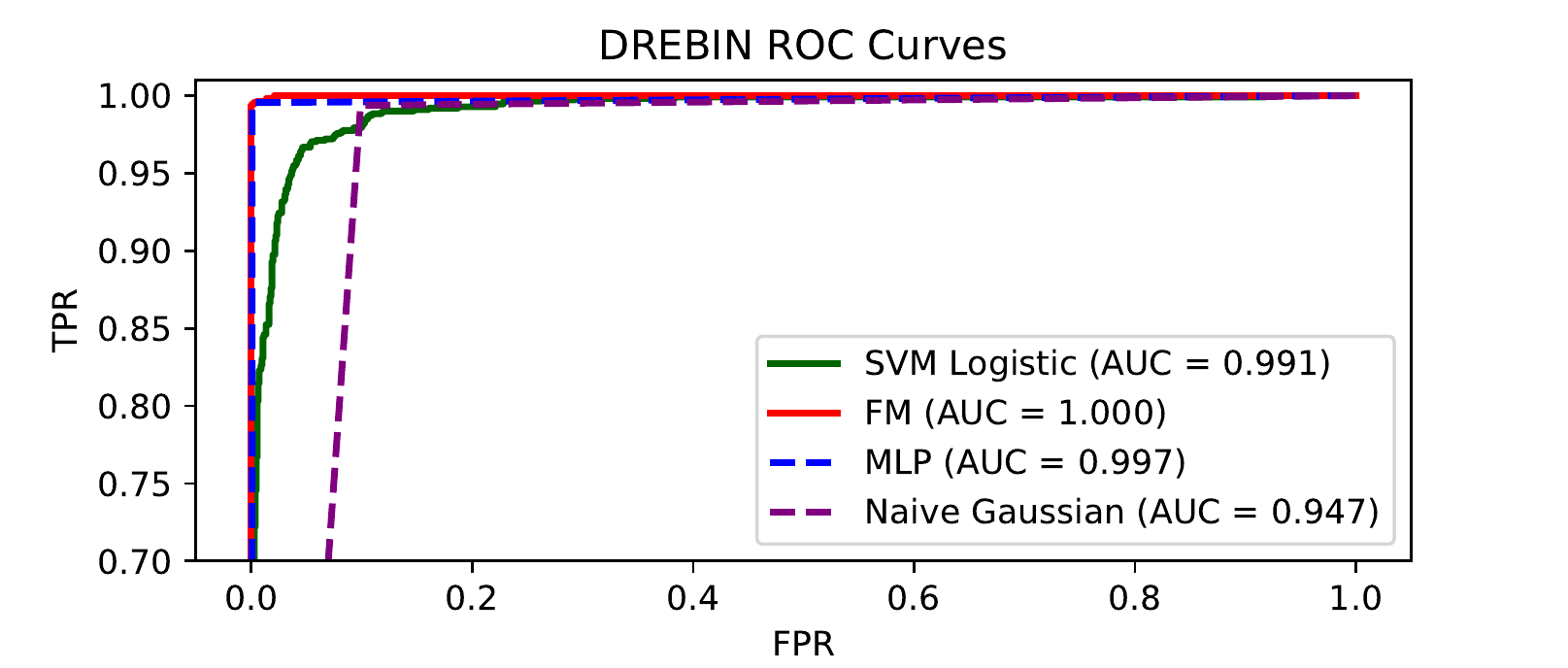}
	\caption{ROC curves for high-performing algorithms on the DREBIN set.}
	\label{fig:drebin_roc}
\end{figure}

\begin{figure}
	\centering
	\includegraphics[width=3.5in]{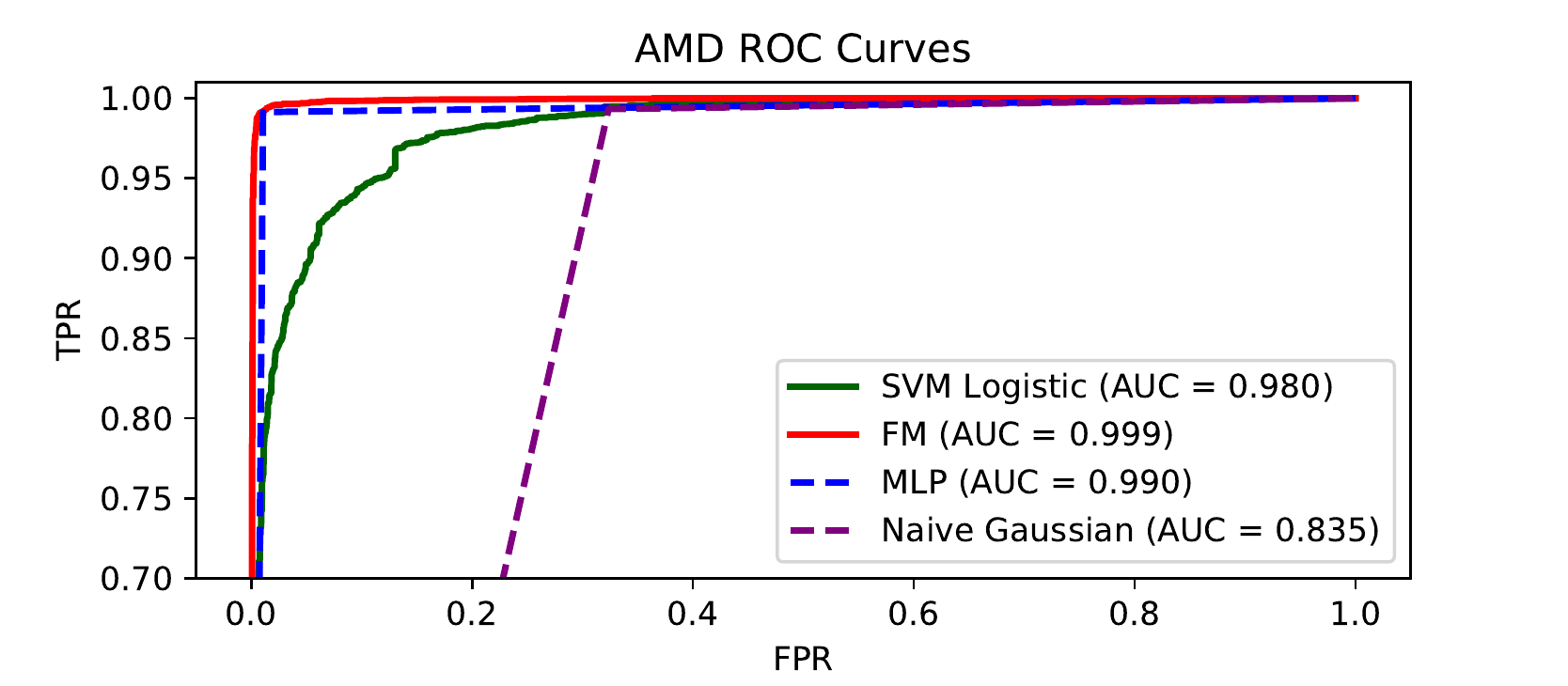}
	\caption{ROC curves for high-performing algorithms on the AMD set.}
	\label{fig:amd_roc}
\end{figure} 

\subsubsection{Interpretation}
\label{ssec:interpretFM_MLP}
Next, Tables~\ref{table:drebin}, \ref{table:amd} and \ref{table:trainTime} illustrate the effectiveness of our FM classifier with respect to SVM with a Logistic Kernel and Gaussian Naive Bayes and especially an MLP. From the first two figures it is clear that the most relevant algorithms are the Mutli-Layer Perceptron and the Factorization Machine. Both classifiers are neck-and-neck with each other; on the DREBIN set the MLP edges out the FM int terms of accuracy with a score of $99.73\%$ to $99.46\%$, however the FM achieved perfect precision and false-positive scores of $100\%$ and $0\%$, respectively. The situation is even tighter when looking at the AMD set, with both classifiers achieving matching accuracy and F1 scores of $99.05\%$ and $99.19\%$, again, respectively. ROC Curves, Figures~\ref{fig:drebin_roc} and \ref{fig:amd_roc} show that FMs pulled ahead of MLPs slightly in terms of area-under-curve (AUC). 

Our Factorization Machine pulled ahead of Support Vector Machines with a Logistic Kernel, verifying our assertion that interation terms are important for revealing malicious behaviour patterns. However, they also cast doubt on our additional assertion that the large number of tunable, synaptic weights used by an MLP would make it hard to train given the sparse data involved. As a universal approximator \cite{hornik1989multilayer, SCARSELLI199815}, MLPs used the same datasets to produce top-notch results at the \emph{cost} of taking much more time to train a classifier. 

Taking a look at Table~\ref{table:trainTime}, we can see that our Factorization Machines trained much faster than our single hidden-layer MLPs. Specifically, the FM trained \textbf{15} times faster than the MLP on DREBIN and \textbf{50} times faster on AMD. This advantage cannot be understated, especially when we consider how much quicker the FM was to train on the AMD set, which was more difficult across the board to get an accuracy score above $99\%$. This emphasis is compounded by the fact that AMD is newer, larger, more recent and therefore relevant than the DREBIN data. Also, note that the MLPs used very basic -- wide and shallow -- and that their training time would only be increased with the additional parameters added by stacking more layers. In short, when compared to Multi-Layer Perceptrons, Factorization Machines trade universal approximation for second-order interactions while striking a balance between time, accuracy, and complexity.

\subsection{Commercial Engines and Family Detection}
\label{ssec:commerical}
We also compared the performance of our malware detection algorithm with existing commercial Anti-Virus engines on VirusTotal \cite{Virustotal2018}. The critical point to mention is that all of the \emph{truly} clean files used in our experiments are actually labeled by these AV engines using the rule described in subsection \ref{ssec:setup}. Therefore, AV engines are supposed to have a better false positive rate than their normal performance. Tables~\ref{table:Virustotaldrebin} and \ref{table:Virustotalamd} summarize the scanning results of the best performing and popular commercial AV engines on \texttt{VirusTotal}, such as \texttt{Kaspersky}, \texttt{Cylance} and \texttt{McAfee} on the DREBIN\cite{arp2014drebin} and AMD\cite{wei2017deep} sets.

To show our model's capacity to distinguish one malware family from other families as well as clean files, we determined whether each input sample belonged to a specific malware family. Here we regard clean files as a special family named ``clean''. We further evaluated our Factorization Machine model for this task on the AMD dataset. Specifically, we used all samples from the $7$ largest malware families in the AMD dataset as well as $1,500$ clean samples.

\subsubsection{Comparison with commercial engines}

\begin{table}[th]
	\centering
	\caption{Test Performance of VirusTotal -- DREBIN}
	\label{table:Virustotaldrebin}
	\begin{tabular}{c|cccc}
		\hline
		Scanner    			& Precision (\%)    & Recall (\%)    & F1 (\%)    & FPR (\%)  \\ \hline
		McAfee        		& \textbf{99.91}  		& 98.74		  & 99.32	& \textbf{0.089}  \\ 
		CAT-QuickHeal		& 99.64			& 99.46		  & 99.55	& 0.357  \\ 
		Symantec  			& \textbf{99.91}			& 99.28		  & \textbf{99.59}	& \textbf{0.089}  \\ 
		Kaspersky  			& 99.63          & 97.21       & 98.41   & 0.357  \\
		Cylance   			& 50.09 			& \textbf{99.91}       & 66.73   & 98.66  \\ 
		Qihoo-360  			& 97.78          & 94.96       & 96.35   & 2.141  \\ 
		\hline
	\end{tabular}
	
\end{table}

\begin{table}[th]
	\centering
	\caption{Test Performance of VirusTotal -- AMD}
	\label{table:Virustotalamd}
	\begin{tabular}{c|cccc}
		\hline
		Scanner    				& Precision (\%)     & Recall (\%)      & F1 (\%)     & FPR (\%) \\ \hline
		McAfee     				& 99.73          & 93.82       & 96.69   & 0.358 \\ 
		CAT-QuickHeal  			& 99.70          & 98.84       & \textbf{99.27 }  & 0.418 \\ 
		Symantec   				& 99.57          & 67.26       & 80.29   & 0.42  \\ 
		Kaspersky  				& \textbf{99.84}          & 53.35       & 69.54   & \textbf{0.119} \\ 
		Cylance    				& 58.86          & \textbf{99.64}       & 74.00   & 98.96 \\ 
		Qihoo-360  				& 97.50          & 68.92       & 80.76   & 2.507 \\ 
		\hline
	\end{tabular}
\end{table}
We can compare our test results in Tables~\ref{table:drebin} and \ref{table:amd} with existing commerical Anti-Virus engines avaliable on VirusTotal \cite{Virustotal2018} in Tables~\ref{table:Virustotaldrebin} and \ref{table:Virustotalamd}. Our results are competitive with the most popular engines listed in the latter two tables, however our best classifiers did fall a little short of the best classifiers available -- \texttt{McAfee}, \texttt{Symantec} and \texttt{Cylance} for DREBIN, \texttt{Kasperskey} and \texttt{Cylance} for AMD. Although we did not record the accuracy of these AV engines, we can see that precision, a metric that measures how many times a classifier accurately deemed a sample to be malware, recall, a measure of how many malware samples the classifier detected in total, and F1, the harmonic mean of both, only fell below $99\%$ for FM on the DREBIN set -- all other times the FM and MLP scores were over $99\%$. The critical point to mention is that all of the \emph{truly} clean files used in our experiments are actually labeled by these AV engines using the rule described in Subsection \ref{exp:datasets}. Therefore, AV engines are supposed to have a better false positive rate than their normal performance. 

\subsubsection{Detection of Specific Malware Families}
\label{ssec:family}

Finally, using Table~\ref{table:familyamdresult}, we can see that the easiest malware to detect was the \emph{Mecor}, which is Trojan Spyware~\cite{wei2017deep}, while the hardest to detect was \emph{Youmi} -- Adware. The brand of malware that was detected with the smallest FPR was \emph{FakeInstaller}, which is a Trojan that wrecks havoc on the device's SMS services. Out of the families listed in in Table~\ref{table:familyamdresult}, the one that is potentially very dangerous yet did not receive one of the highest scores was \textit{Fusob}~\cite{wei2017deep}, ransomware which can lock down the device until certain conditions, which usually involve monetary payment to the hacker in question, are met. However, our FM design scored over $99\%$ in the fields of precision, recall and F1 on this scarce family of $1,238$ entries.  Ironically, our classifier had a harder time deeming \texttt{apk} files to be clean than it did detecting any brand of malware, but as the saying goes better safe than sorry. 

\begin{table}[]
	\centering
	\caption{Malware Family Classification Results by FM -- AMD}
	\label{table:familyamdresult}
	\begin{tabular}{l|ccccc}
		\hline
		Family          & Samples   & Precision (\%)   & Recall (\%)  & F1 (\%)    & FPR (\%)       \\ \hline
		Airpush       & 7606  & 99.54          & 99.72          & 99.63          & 0.17          \\
		Youmi         & 1256  & \textit{97.53}          & \textit{98.75}          & \textit{98.14}          & 0.09          \\
		Mecor         & 1762  & \textbf{99.77} & \textbf{99.89} & \textbf{99.83} & 0.11          \\
		FakeIns. 	  & 2129  & 99.57          & 99.57          & 99.57          & \textbf{0.05} \\
		Fusob         & 1238  & 99.68          & 99.52          & 99.60          & \textit{0.48 }         \\
		Kuguo         & 1122  & 99.64          & 99.82          & 99.73          & 0.18          \\
		Dowgin        & 3298  & 98.52          & 99.63          & 99.07          & 0.07          \\
		Clean         & 1500  & 95.58          & 97.30          & 96.43          & 0.22          \\
		Average       & ---   & 98.73          & 99.27          & 99.00          & 0.17          \\
		\hline
	\end{tabular}
\end{table}

\subsection{Feature Processing Overhead}
\label{ssec:process}

\begin{figure}
	\centering
	\includegraphics[width=3in]{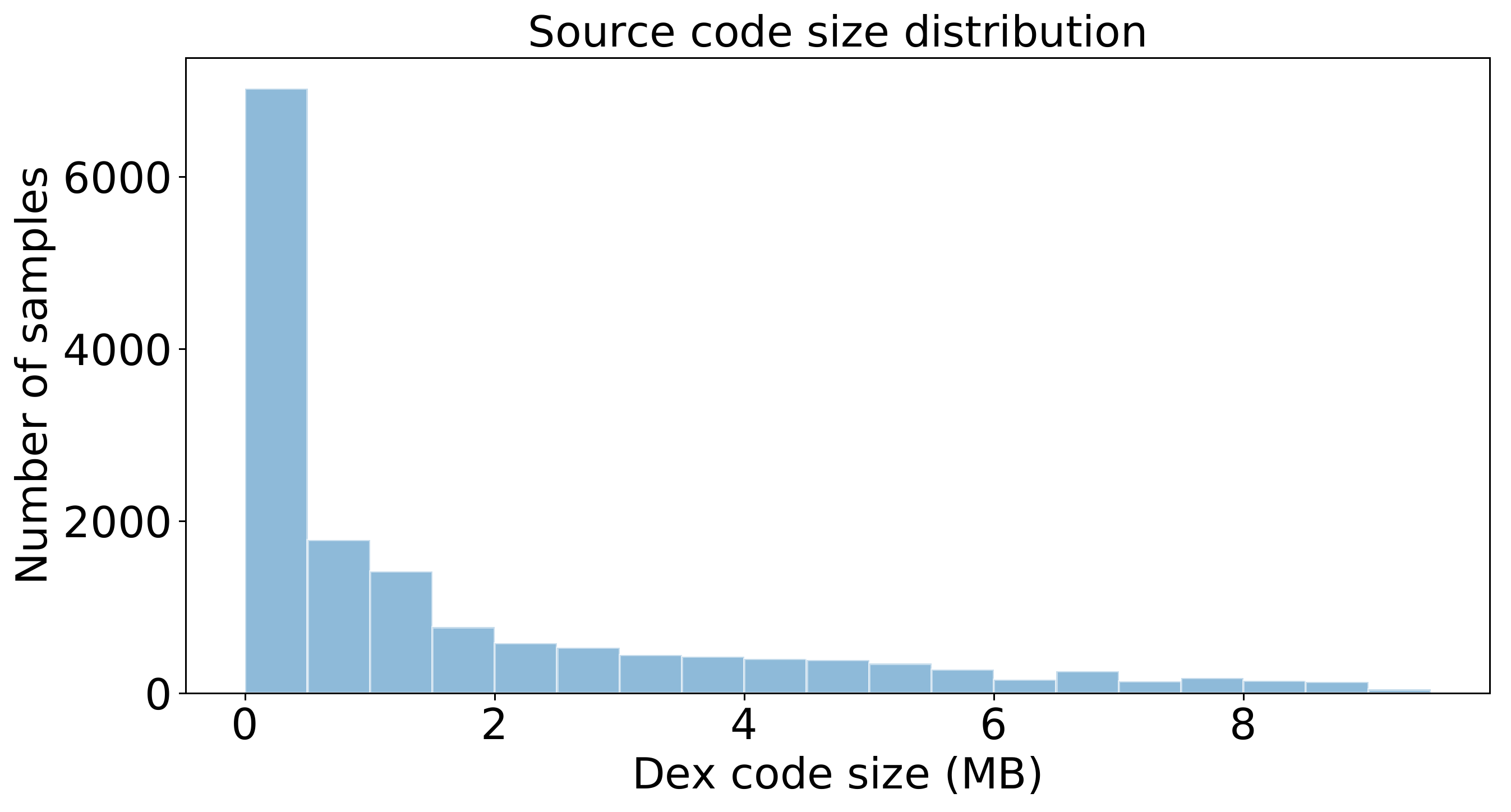}
	\caption{Dex source code size distribution.}
	\label{fig:codesize}
\end{figure}

\begin{figure}
	\centering
	\includegraphics[width=3in]{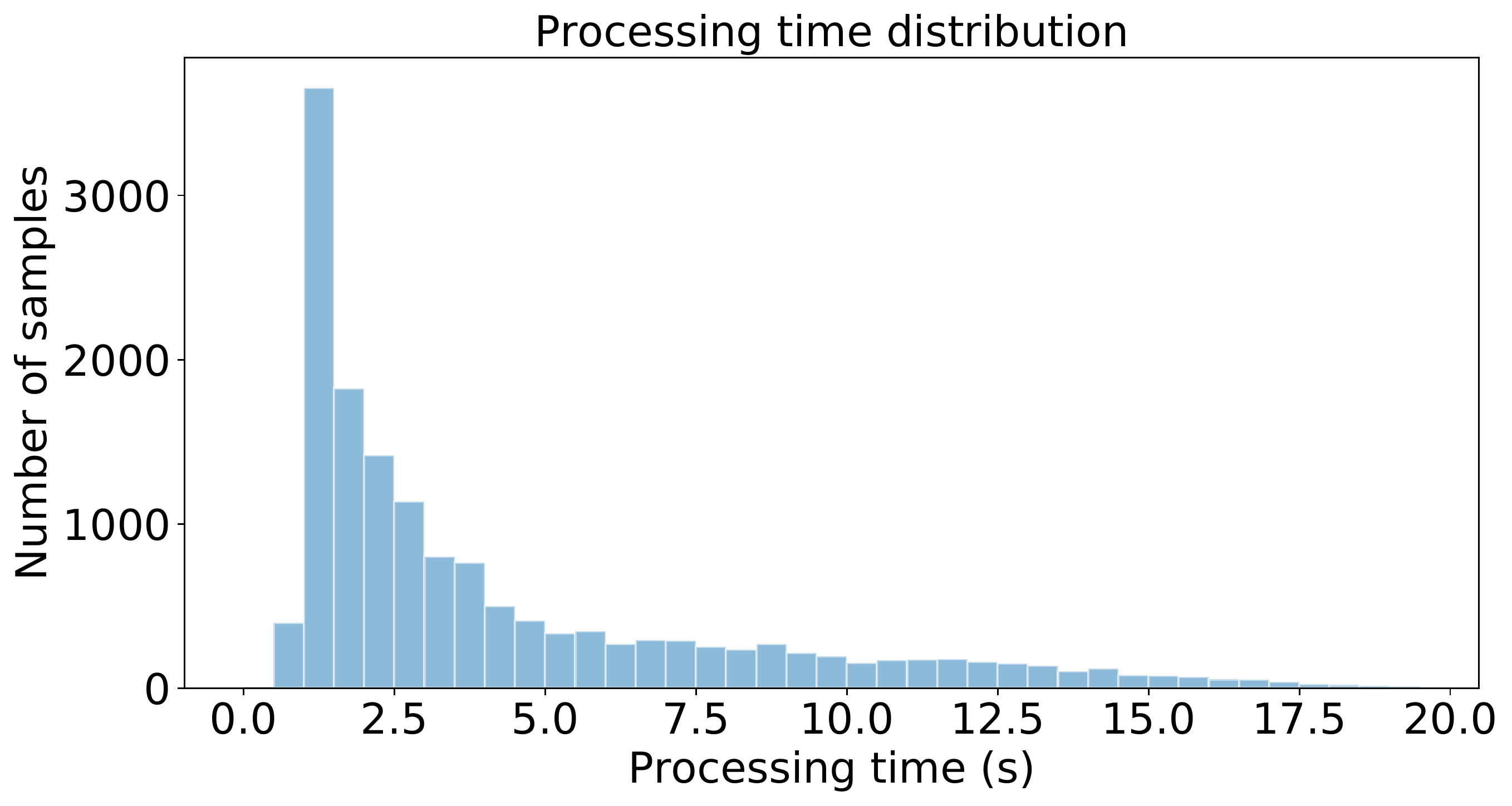}
	\caption{Processing time distribution.}
	\label{fig:time}
\end{figure}

\begin{figure*}[ht]
  \centering
  \subfigure{
    \includegraphics[height=1.5in]{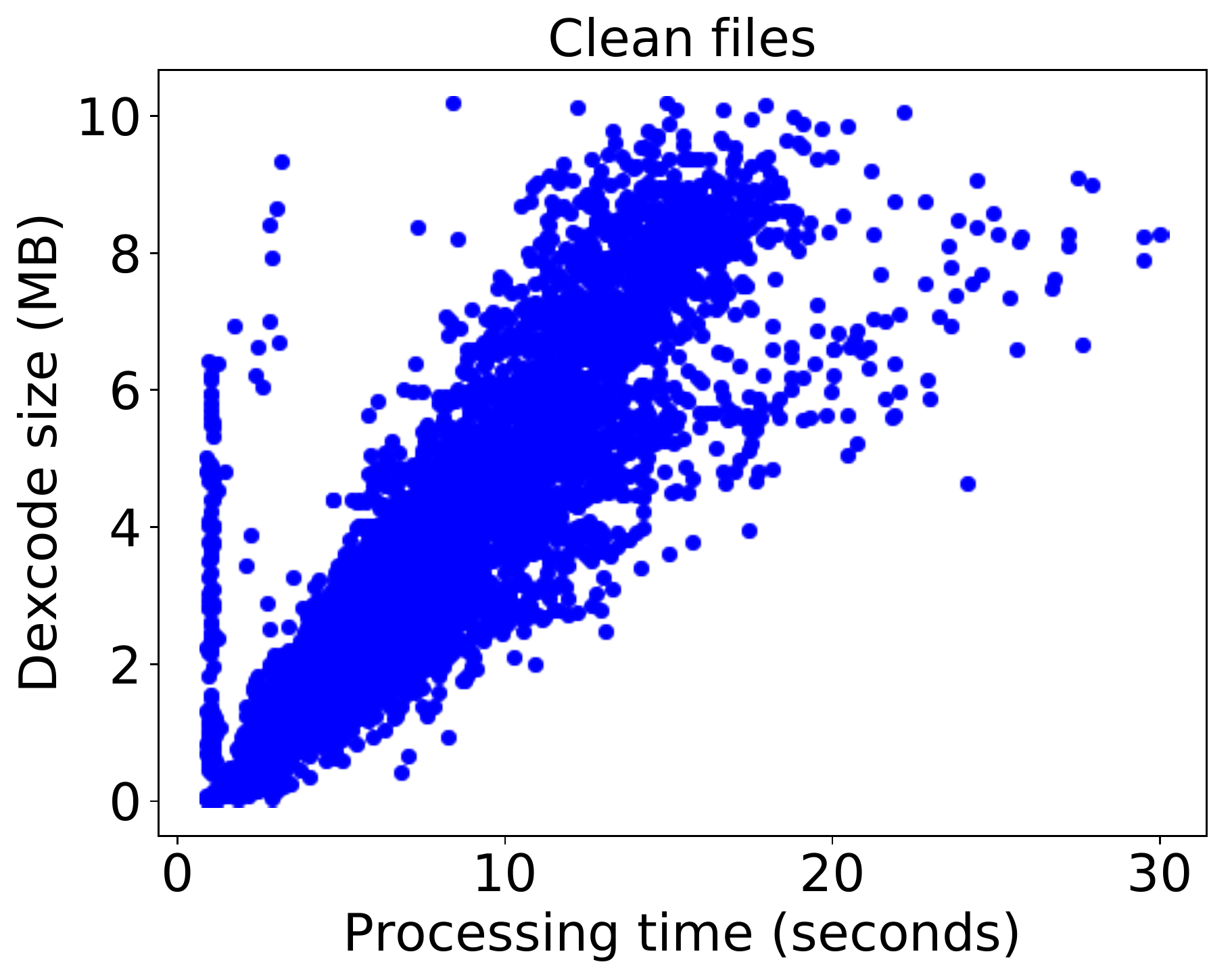}
  }
  \hspace{3mm}
  \subfigure{
    \includegraphics[height=1.5in]{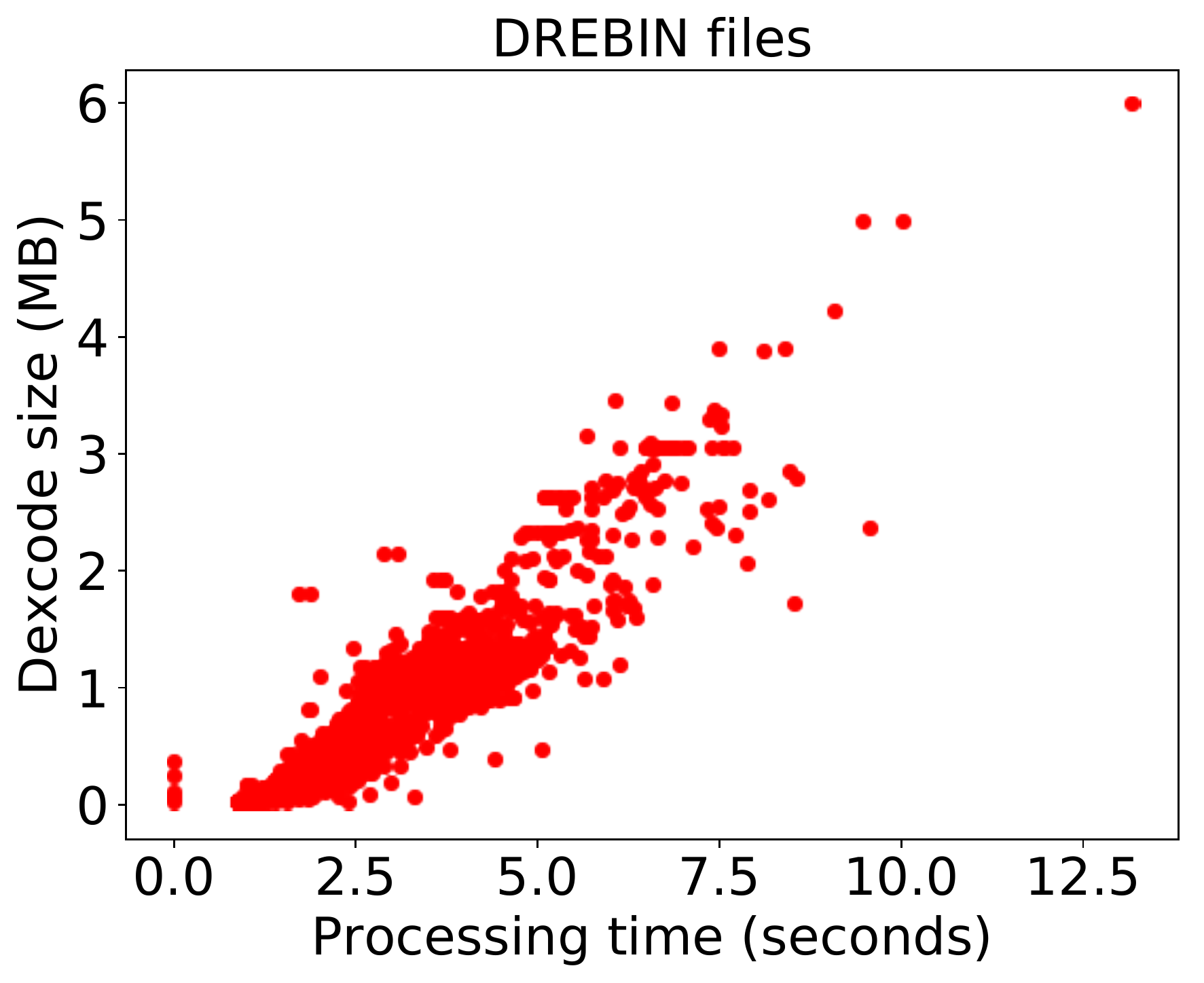}
  }
  \hspace{3mm}
  \subfigure{
    \includegraphics[height=1.5in]{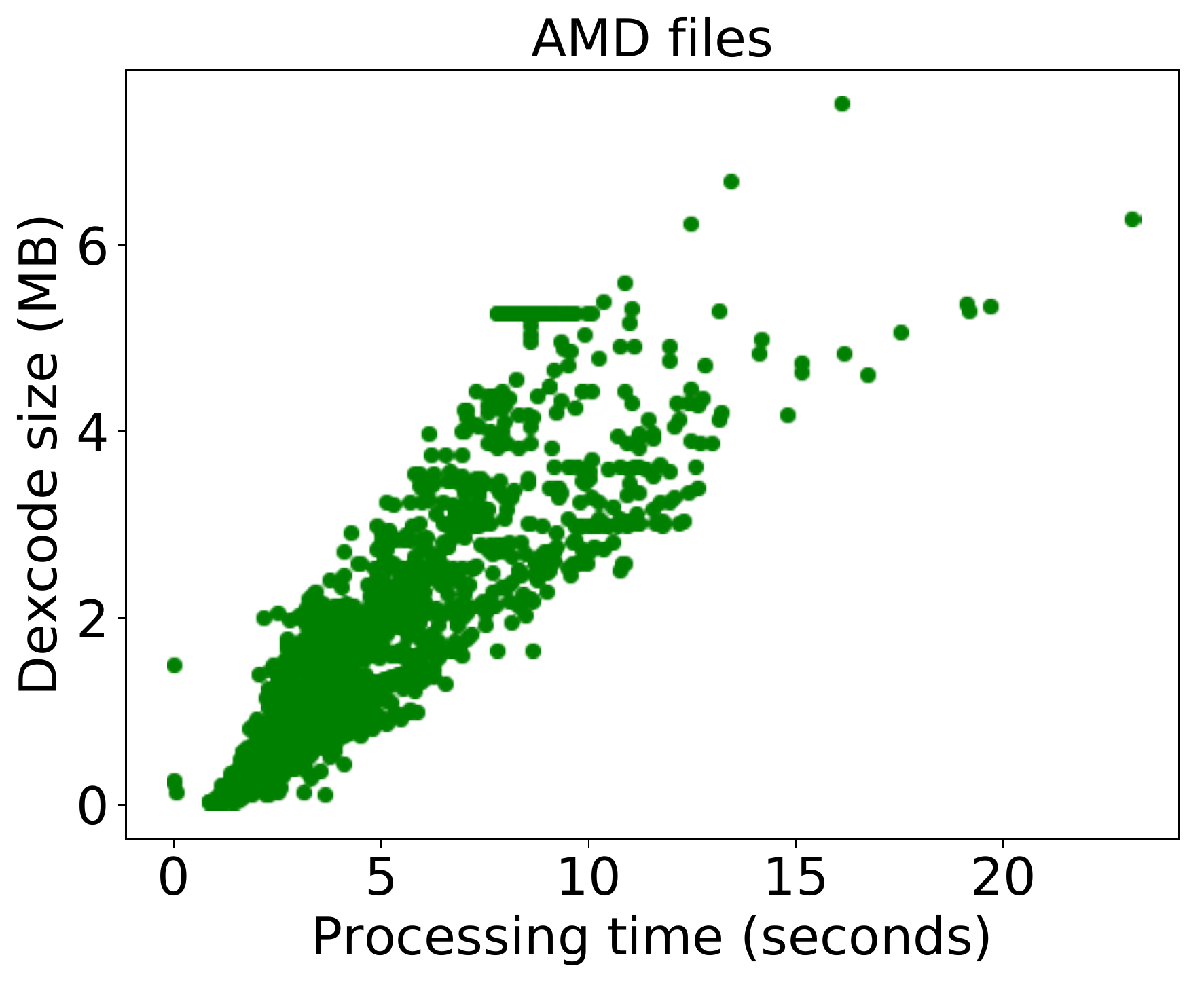}
  }
  \vspace{-3mm}
  \subfigure{
    \includegraphics[height=1.5in]{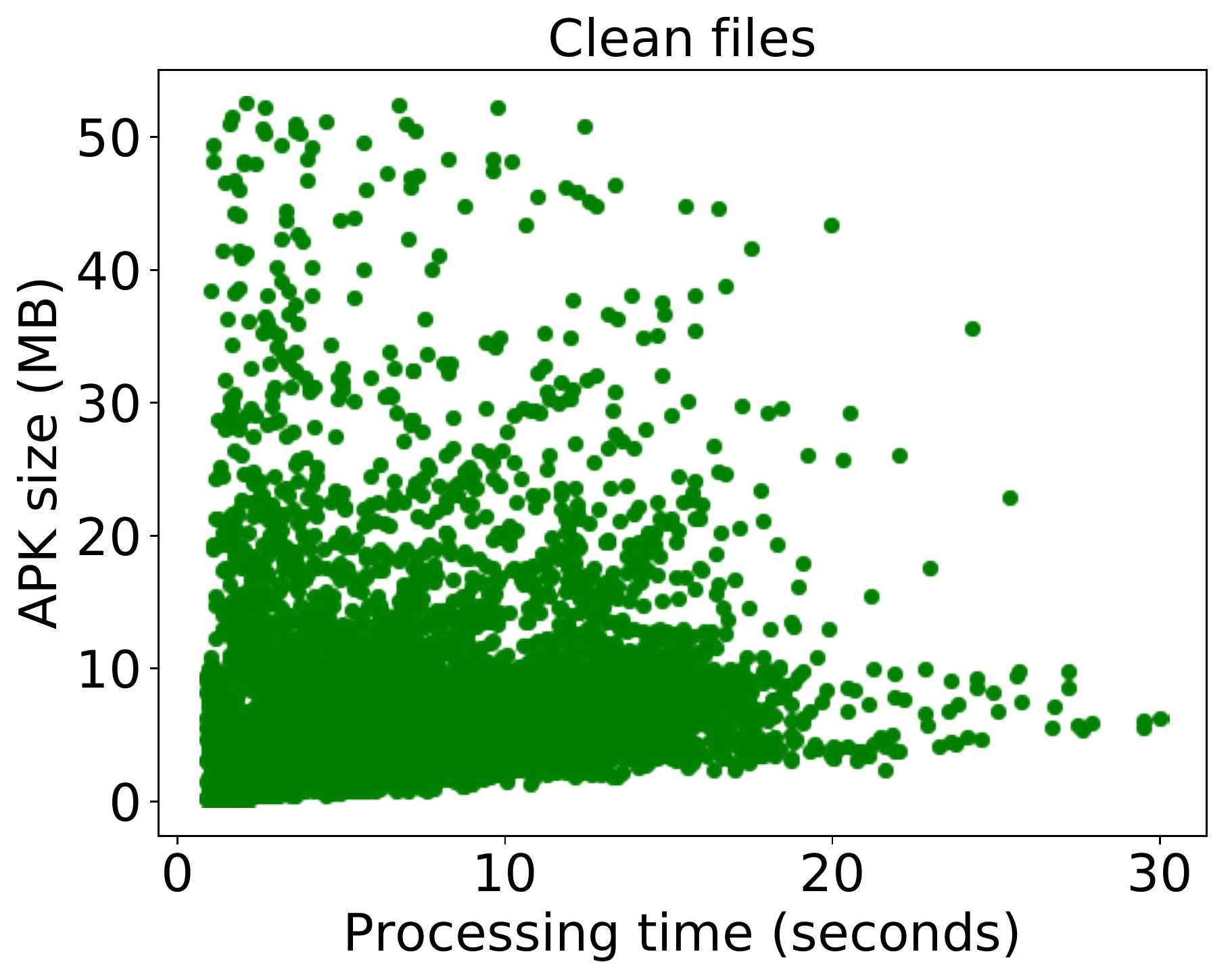}
  }
  \hspace{3mm}
  \subfigure{
    \includegraphics[height=1.5in]{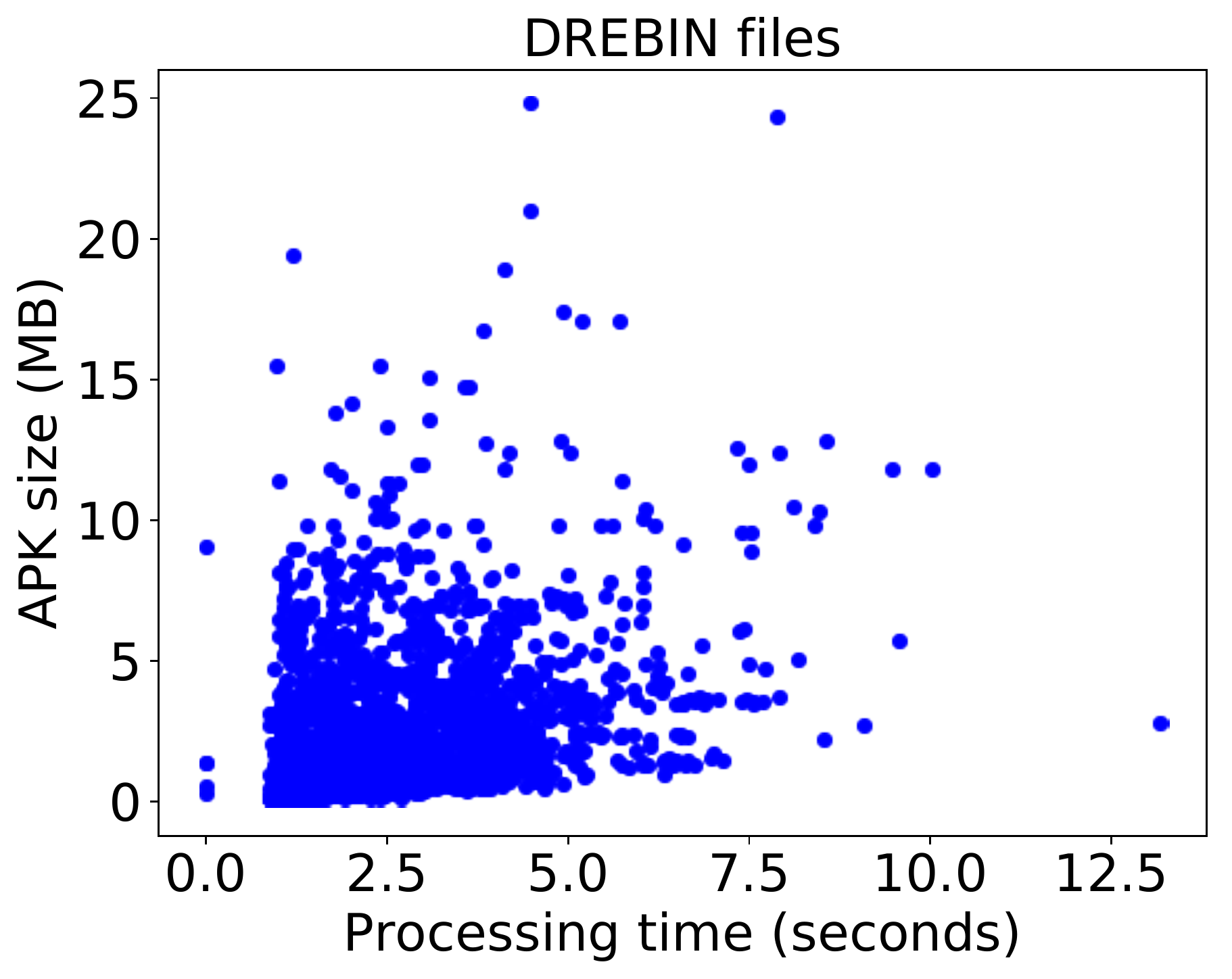}
  }
  \hspace{3mm}
  \subfigure{
    \includegraphics[height=1.5in]{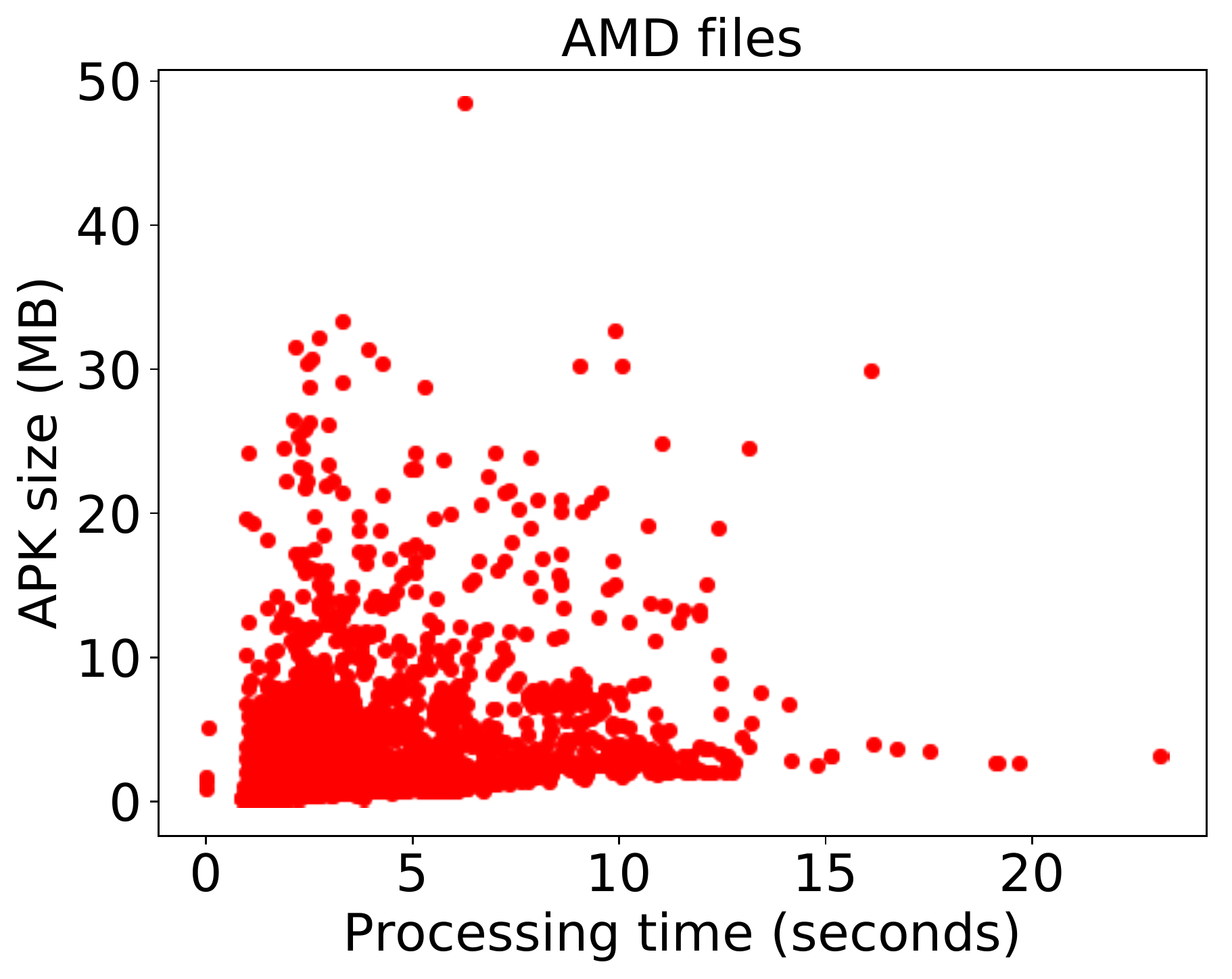}
  }
	\caption{Scatter plot of processing time vs. file size.}
	\label{fig:runtime}
\end{figure*}

Now we evaluate pre-processing time which consists of decompiling the \texttt{apk} files to Smali code and then extracting the features listed in Section~\ref{sec:prelim}. All work relating to this subsection was done on a virtual machine hosted on ESXi. The VM was running Ubuntu 16.04 with a memory of $4$G and 2 CPUs. For this task we randomly sampled $3,794$ AMD samples, $6,120$ clean files and all $5,560$ DREBIN samples. Histograms for dex code size and processing time for all $15,474$ samples are given in Figures~\ref{fig:codesize} and \ref{fig:time}, respectively. The results of processing time vs. size-on-disk are shown in Figure~\ref{fig:runtime}; the three figures in the first row show the relation between dex source code size and processing time. The figures in the second row show the relation between \texttt{apk} file size and processing time.



\subsubsection{Discussion of Results}
\label{ssec:discuss}

With respect to \texttt{apk} file decompiling and feature extraction, Figures~\ref{fig:codesize}, \ref{fig:time} show us that over $78\%$ of samples have a dex code size of less than $3MB$ and over $70.6\%$ samples have a processing time of less than $5$ seconds. On the same samples we also measured the mean time for encoding and prediction. Figure \ref{fig:runtime} tells us that the relationship between processing time and dex code size is almost linear and that for the samples in all three datasets the slopes are in the neighborhood of $0.4$. Conversely, the three top graphs of Figure~\ref{fig:runtime} indicate no fixed relation between \texttt{apk} file size and processing time, but rather that the relationship between \texttt{apk} size and processing time has a rough upper and lower bound. It is likely that this decoupling of processing time between \texttt{apk} file size and dex code size is because in addition to the dex code and manifest file, an \texttt{apk} file also contains other resource files like HTML, figures, which can vary wildly from app-to-app.  

Compared with DREBIN \cite{arp2014drebin}, it seems that our system does not have much of an advantage in processing time. However, this is not the case. To begin with, the test is done on a system that is not fully integrated, the output of Smalisca is first written into a json file and then reload into RAM for further processing. The I/O between RAM and flash storage would often take a long time. Secondly, the feature sets used in our system are simpler and smaller than sets used in DREBIN, so under same condition our system should take less processing time than DREBIN.

\section{Limitations and Future Work}
\label{sec:future}

While machine learning techniques such as ours provide a powerful tool for automatically inferring models, they require a representative dataset for training. That is, the quality of the detection model depends on the availability, quality and quality of both malware and benign applications. While it is straightforward to collect benign applications, gathering recent malware samples is a non-trivial effort that requires some technical effort. 
Fortunately, offline analysis methods, e.g. RiskRanker \cite{grace2012riskranker}, can help to acquire malware and provide the samples for updating and maintaining a representative dataset in order to continuously update our model.


Outside of model training times, the major limitation of our architecture is the decompilation and feature extraction process. We plan to integrate our system into Wedge Networks' in-line, real-time security solution which only allows us to have millisecond-scale processing time. For encoding and prediction our system takes about $4.8$ms on average, however, decompiling and feature extraction is on the order of seconds. Fortunately, methods exist that will allow us to improve our system's time efficiency. Such techniques include reducing I/O and finishing all work at once on a computer with a large amount of main memory (RAM), or even using Application-Specific Integrated Circuits (ASIC) such as FPGAs for speed up. In addition, we noted that decompiling \texttt{apk} files can fail when using some existing tools. In our experiments, we observed failures for some files, more with malware samples than clean files. This is expected as malware samples may use some additional techniques such as code obfuscation~\cite{malwareObfuscation} that may lead to decompiling failures. This limits the effectiveness of Android malware detection schemes that extract features from \texttt{apk} files and serves as an avenue for future research.


\section{Related Work}
\label{sec:related}




Many recent papers are trying to find malicious behavior patterns through control flow graphs or call graphs, although these can be obfuscated by "method overloading"~\cite{malwareObfuscation}. AppContext \cite{yang2015appcontext} classifies applications using machine learning based on the contexts that trigger security-sensitive behaviors. It builds a call graph from an application source code and extracts the context factors through information flow analysis. It is then able to obtain the features for the machine learning algorithms from the extracted context. In that paper, 633 benign applications from the Google Play store and 202 malicious samples were analyzed. AppContext correctly identifies 192 of the malware applications with an $87.7\%$ accuracy. \citet{gascon2013structural} also utilized call graphs to detect malware. After extraction of call graphs from Android applications, a linear-time graph kernel is applied in order to map call graphs to features. These features are given as input to SVMs to distinguish between benign and malicious applications. They conducted experiments on $135,792$ benign and $12,158$ malware applications, detecting $89\%$ of the malware with an FPR of $1\%$. This kind of method relies heavily on the accuracy of call graph extraction. However, current works like FlowDroid \cite{arzt2014flowdroid} and IC3 \cite{octeau2015composite} cannot fully solve the construction of Inter-component control flow graphs (ICFG), especially the inter-component links with intents and intent filters.

Other works focus on the detection of specific malicious behavior such as privacy breaches and over privilege usage. For example, \cite{kim2012scandal} goes through the source code with predefined sources and sinks to find a potential privacy breaches. 
\cite{fu2017leaksemantic} further examines all the URL addresses to see if the app is trying to steal users' private information. 
\cite{fuchs2009scandroid} uses data flow analysis for security certification. However, static taint-analysis and over privilege are prone to false positives.

Studies closer to the one performed in this paper, such as \cite{hou2017hindroid,peiravian2013machine} try to directly classify an application as malicious or benign through permission request analysis for application installation \cite{felt2012android}, or control flow analysis \cite{liang2014fast}. 
These works take different approaches in both the feature extraction and the classification phase. \citet{peiravian2013machine} used permissions and API calls as features for SVM and Decision Tree Ensemble classifiers. 
Hindroid \cite{hou2017hindroid} built a structured heterogeneous information network (HIN) with an Android application and related system APIs as nodes and their rich relationships as links, and then used meta-paths for malware detection. DREBIN \cite{arp2014drebin}, which extracted features from manifest files and source code, including permissions, hardware, system API calls and even all the URLs, and then used SVM as the final classifier for malware detection. However, DREBIN only achieved a test detection rate of $93.90\%$ on their full dataset. \citet{sahs2012machine} used a one-class SVM with kernels and as general classifier, but only used 2081 benign and 91 malicious applications. Next, \citet{yuan2014droid} used Deep Neural Networks for malware detection, but they restricted themselves to only 500 \texttt{apk} files and achieved an test accuracy of $96.5\%$. Finally, in an extensive study that tested the ability of past malware's effectiveness at training classifiers to deal with new malware, \citet{OnwuzurikeMarkovChains} used a mix of call graphs and markov chains for prediction however their metrics focused on the F-measure\footnote{Generalization of the F1 score} and their results greatly varied depending on across different sets of data. By contrast the accuracy scores of our single hidden-layer MLP and FM models, were each above $99\%$ on both datasets.

We believe one of the reasons behind these high scores is that we differentiated ourselves from existing works and instead of only focusing on feature engineering and ignoring the importance of choosing a suitable algorithm, after acquiring the feature representations of apps, we first made two critical observation regarding the interaction between features and the sparsity of the feature vectors. Then, the optimum machine learning algorithm designed to appropriately handle our problem was chosen for malware detection based on those observations and assumptions.

\section{Conclusion}
\label{sec:conclude}


In this paper, we raised the issue of considering interaction terms across features for the discovery of malicious behavior patterns in Android applications. The features used to represent an \texttt{apk} file consisted of app components, hardware features, permissions, intent filters from the manifest file, restricted APIs, suspicious APIs and used permissions from source code. Based on the extracted features, a highly sparse vector representation was constructed for each application using one-hot encoding. We then proposed the use of a Factorization Machine-based malware detection system to handle the high sparsity of vector representation and model interaction terms at the same time.

To the best of our knowledge, this is a first for using FM models for malware detection. A comprehensive experimental study on two real sample malware collections, the DREBIN and AMD datasets, alongside clean applications collected from online app stores were performed to show the effectiveness of our system on malware detection and malware family identification tasks. Promising experimental results with accuracy, precision, recall and F1 scores of around or above $99\%$ demonstrated that our method matches the performance of commercial antivirus engines and holds steady against the incredible results produced by Multi-Layer Perceptrons with the benefit of taking up to 50 times less time to train.

\bibliographystyle{ACM-Reference-Format}
\bibliography{main}

\end{document}